\documentclass[double]{article}
\usepackage{times}
\usepackage{algorithm}
\usepackage{algorithmic}
\usepackage{wrapfig}
\usepackage{verbatim}
\usepackage{amsmath}
\usepackage{graphicx}
\usepackage{subcaption}
\usepackage[active]{srcltx}
\usepackage{color}
\usepackage{lineno}
\usepackage{dirtytalk}
\usepackage{amsthm}
\usepackage{authblk}
\usepackage{listings}
\usepackage{tikz}
\usepackage{longtable}
\usetikzlibrary{shapes.geometric, arrows}

\usepackage{epsfig,graphicx,amsfonts}
\usepackage{amsmath,amsthm,amssymb}
\usepackage{xcolor}

\usepackage{adjustbox}
\usepackage{multirow}
\usepackage{setspace}
\usepackage{hyperref}
\usepackage{comment}

\long\def\comment #1\commentend{}

\begin{document}

\title{\Large Spatio-Temporal SIR Model of Pandemic Spread During Warfare with Optimal Dual-use Healthcare System Administration using Deep Reinforcement Learning}

\author{Adi Shuchami$^{1}$, Teddy Lazebnik$^{1,2}$\\
\(^1\) Department of Mathematics, Ariel University, Ariel, Israel\\
\(^2\) Department of Cancer Biology, Cancer Institute, University College London, London, UK\\
\(^*\) Corresponding author: nogal@ariel.ac.il
}

\date{ }

\maketitle 

\begin{abstract}
\noindent
Large-scale crises, including wars and pandemics, have repeatedly shaped human history, and their simultaneous occurrence presents profound challenges to societies. Understanding the dynamics of epidemic spread during warfare is essential for developing effective containment strategies in complex conflict zones. While research has explored epidemic models in various settings, the impact of warfare on epidemic dynamics remains underexplored. In this study, we proposed a novel mathematical model that integrates the epidemiological SIR (susceptible-infected-recovered) model with the war dynamics Lanchester model to explore the dual influence of war and pandemic on a population's mortality. Moreover, we consider a dual-use military and civil healthcare system that aims to reduce the overall mortality rate which can use different administration policies. Using an agent-based simulation to generate \textit{in silico} data, we trained a deep reinforcement learning model for healthcare administration policy and conducted an intensive investigation on its performance. Our results show that a pandemic during war conduces chaotic dynamics where the healthcare system should either prioritize war-injured soldiers or pandemic-infected civilians based on the immediate amount of mortality from each option, ignoring long-term objectives. Our findings highlight the importance of integrating conflict-related factors into epidemic modeling to enhance preparedness and response strategies in conflict-affected areas. \\ \\

\end{abstract}

\maketitle \thispagestyle{empty}
\pagestyle{myheadings} \markboth{Draft:  \today}{Draft:  \today}
\setcounter{page}{1}

\section{Introduction}
\label{sec:introduction}
Throughout history, pandemics have significantly influenced societal structures, economic stability, political systems, and impacted mortality rates
\cite{pandemic_important}. In recent decades, the ongoing threat of emerging and reemerging pandemics and epidemics has posed a significant challenge to humanity \cite{pandemic_duration}. In a similar yet distinct manner, the overall number of wars and conflicts worldwide has shown a concerning increase recently, following a relatively peaceful period after World War II \cite{more_wars_1,more_wars_2}. 

Pandemics and military operations have long been interlinked due to the sociological and operational dynamics associated with armies and warfare \cite{pandemic_war_intro_2,pandemic_war_intro_1}. The levels of hygiene, self-care, and immune system stress in individuals provided fertile conditions for the outbreak of diseases within armies and the communities interacting with them \cite{army_camp_pandemic}. Moreover, the density of military bases and army deployments facilitates the spread of pathogens within the population \cite{army_camp_pandemic}. Notable examples include the 1917-1918 Spanish Influenza, which affected populations worldwide and particularly impacted the United States (US) military during World War I, and the COVID-19 outbreak in Ukraine during the Russia-Ukraine war starting in 2022 \cite{spanish_1918_ww1,covid_russia_war_2022}. 

The intersection of pandemics and military operations has placed economic and healthcare systems worldwide under considerable strain \cite{edge_1,edge_2,edge_3}. To prepare for such events, researchers frequently design policies and plans for a wide range of scenarios using mathematical models \cite{model_1,model_2,model_3,model_4,model_5}. Models related to warfare are often less accessible to the general public and, consequently, can be difficult to locate in the literature \cite{war_models_not_alot}. Unlike, numerous models have been proposed to simulate the spread of pandemics \cite{model_sir_review_1,model_sir_review_2,lazebnik2021novel}. For example, \cite{exp_2_2_1} proposed a SIR (Susceptible-Infected-Recovered) based disease spread model that incorporates multiple transmission pathways, including direct contact and vectors. This model effectively captures the dynamics of pandemic spread across large populations. \cite{clinical_2} introduced an extended SIR model that differentiates between symptomatic and asymptomatic cases of COVID-19, assigning distinct infection paths for each. In this model, susceptible individuals can become asymptomatic immediately upon infection, while those who develop symptoms first enter an exposed stage. \cite{exp_2_2_3} proposed an extended spatio-temporal SIR model for the spread of sexually transmitted diseases on a bipartite random contact network.

Building on top of these modeling efforts, a growing body of work has focused on the capacity of healthcare systems to support the treatment of infected individuals during large-scale pandemics \cite{healthcare_pandemic}. For instance, \cite{clinical_3} utilized a SIR model incorporating the dynamics of hospitalized patients, assumed to be proportional to the number of infected individuals, with considerations for delay and exponential decay during their stay in healthcare centers. Notably, the hospitalized population does not affect other epidemiological states, which is crucial for accurately modeling healthcare center dynamics. This model was applied to COVID-19 pandemic data from the first three months in five different regions: Belgium, France, Italy, Switzerland, and New York City (US). Similarly, \cite{clinical_1} introduced a risk-stratified SIR-HCD model that incorporates variables representing the dynamics of low- and high-contact subpopulations, with shared parameters to manage their respective rates over time. The authors tested their model using data on daily reported hospitalizations and cumulative mortality from COVID-19 in Harris County, Texas.

\begin{figure}[!ht]
    \centering
    \includegraphics[width=0.99\textwidth]{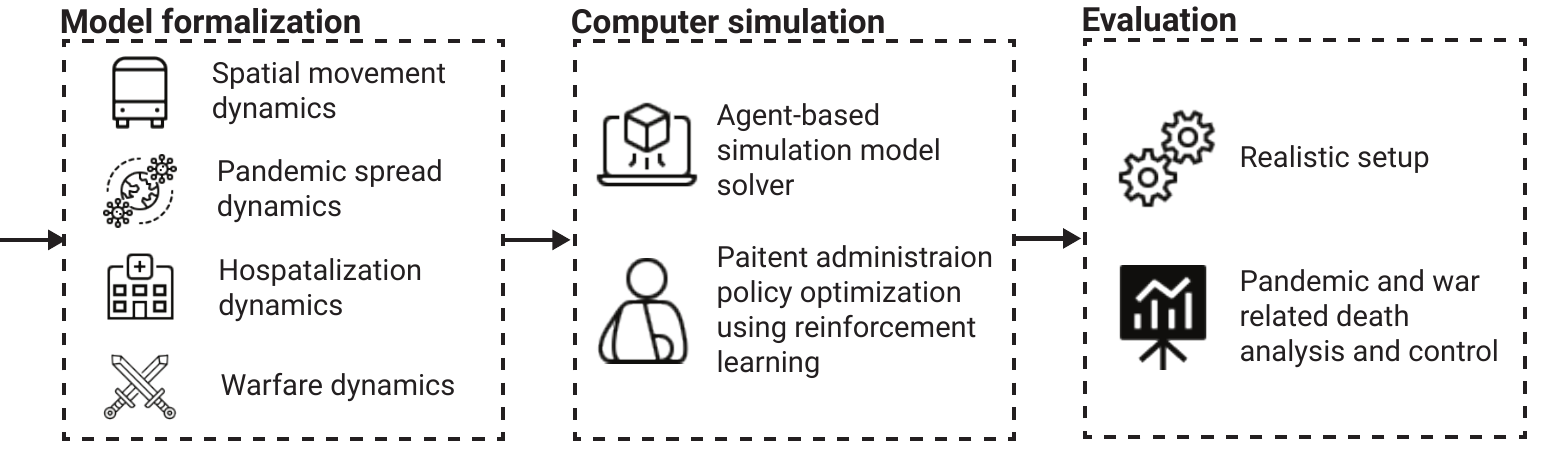}
    \caption{A schematic view of the study's design. First, we formalize the model by introducing four key components: spatial movement dynamics, pandemic spread dynamics, hospitalization dynamics, and warfare dynamics. Next, we utilize agent-based simulation to solve the model and optimize patient administration policies using reinforcement learning. Finally, we establish a realistic setup to evaluate the pandemic- and war-related deaths, analyzing the impact of various patient administration strategies on these outcomes. }
    \label{fig:study_scheme}
\end{figure}

Despite the extensive epidemiological-mathematical research, a model that captures the dynamics of a pandemic during wartime and its influence on the healthcare system's ability to provide necessary services has yet to be developed. This gap leaves policymakers without crucial insights for planning in such scenarios. Hence, in this work, we propose a novel spatio-temporal extended SIR-based model that includes civilian and military healthcare systems with asymmetric administration policies to study the effect of pandemics on warfare performance from the healthcare perspective. In addition, we provide an agent-based simulation implementation of the proposed model with a reinforcement learning model that is able to approximate optimal patient administration policies for the entire healthcare system, including both war and civilian zones to reduce overall mortality (from both war and pandemic), operating as an applicative tool to investigate efficient administration polices with different properties without the need to manually search for them. A schematic view of the study's structure is present in Fig. \ref{fig:study_scheme}. 

The remaining paper is organized as follows: Section \ref{sec:related_work} presents the related work for pandemic spread during wars and spatio-temporal modeling practices for pandemic spread. Section \ref{sec:model} describes the proposed model's mathematical formalization. Section \ref{sec:simulator} outlines the implementation of the proposed model as a computer simulator and optimal patient administration using a reinforcement learning model. Section \ref{sec:analysis_numerical} provides a comprehensive evaluation of the proposed model. Finally, Section \ref{sec:discussion} provides a conclusion on the possible applications and limitations followed by suggestions for future work.  

\section{Related Work}
\label{sec:related_work}
Warfare has been studied from various perspectives, including sociological \cite{war_socio}, political \cite{war_political}, and economic \cite{war_economy}, among others. In addition, over the last century, advancements in biology, medicine, and engineering have greatly enhanced our epidemiological understanding \cite{epi_more_knowledge}. As such, the relationship between the two started to emerge in the literature and a growing body of work captured the interplay between warfare and pandemics. In this section, we review the unique relationship between warfare and pandemic spread. Furthermore, we present recent modeling practices for pandemic spread which will be the mathematical base for our model.

\subsection{Pandemic during war}
In the context of large-scale pandemics and wars, one can detect two central and relatively recent global-scale events - the Spanish influenza pandemic of 1917-1918 during the First World War \cite{spanish_1918_ww1} and the COVID-19 pandemic during the Russian-Ukrainian war \cite{covid_russia_war_2022}. 

Following the sequence of events, we first examine the Spanish influenza pandemic, which stands out as one of the deadliest pandemics in recent history \cite{influnza_pandemic_worst}. Estimates place the global death toll between 21.5 million and 39.3 million, with some scholars suggesting even higher numbers, though these may be exaggerated \cite{spanish_1918_death_count}. The initial wave of the pandemic, occurring in the spring of 1918, was relatively mild with few fatalities. However, a drastic change occurred in the summer of the same year when the virus underwent a mutation, becoming highly virulent and leading to millions of deaths worldwide in October and November  \cite{spanish_1918_history}. A less severe third wave followed in early 1919, while the fourth and final wave extended into the initial months of 1920. The demographic most affected comprised young, healthy adults aged 15 to 44. Mortality rates varied across countries and continents, with estimates in Europe ranging from 1.1\% to 1.2\% \cite{spanish_1918_history_2}. Importantly, \cite{spanish_1918_source} proposed that the pandemic originated at the British military base in Étaples, located in northern France in the Pas-de-Calais department. Significant during World War I, this base housed 100,000 soldiers within a 12 square kilometer area. Adjacent to the base were coastal marshes frequented by migratory birds. In the vicinity, numerous farms kept pigs, ducks, and geese for the soldiers' consumption, along with horses used for transportation. The convergence of dense soldier populations, various livestock, and the extensive use of 24 types of war gases, several of which were mutagenic, could potentially explain the initial outbreak of the epidemic between December 1916 and March 1917. Moreover, \cite{spanish_1918_ww1} suggested that the Spanish influenza was exported by boat to the USA from Tanger and the Spanish colonies in the north of Africa, highlighting the influence of military personnel movement as a vector of pandemic spread.

The Russian-Ukraine war began in February 2022, during the third year of the COVID-19 pandemic. By that time, the pandemic, caused by the SARS-CoV-2 pathogen, had resulted in over six million deaths \cite{covid_russia_intro}. Before the pandemic, Ukraine's healthcare system had many shortcomings, making it difficult to detect and control the spread of COVID-19 \cite{covid_russia_intro}. These challenges included a lack of essential healthcare equipment, inadequate personal protective gear, a shortage of skilled healthcare workers, limited availability of infectious disease departments across different regions of Ukraine, insufficient surveillance measures, and a scarcity of authorized testing laboratories. These issues indicate that Ukraine was already struggling with the COVID-19 crisis, a situation made worse by the ongoing conflict \cite{covid_russia_intro}. Due to the Russian invasion of Ukraine, approximately five million Ukrainians immigrated to neighboring European countries, such as Poland, Romania, Russia, Hungary, Moldova, Slovakia, and Belarus. The invasion also caused population mixing inside of the country as people relocated from war zones \cite{russia_ukrina_war_pandemic}. During the invasion, the Ukrainian community experienced major disruptions in primary healthcare services and faced challenges accessing basic preventive medicine \cite{covid_uk_problem}. Additionally, COVID-19 vaccination efforts and routine immunizations nearly ceased nationwide. The rise in COVID-19 cases in Ukraine caused concern for the European Union. The large influx of refugees—the largest since World War II—posed a risk of spreading the virus beyond Ukraine's borders, potentially triggering new outbreaks across Europe \cite{covid_eu}. 

\subsection{Spatio-temporal pandemic modeling}
Pandemics can arise from various sources, including sexually transmitted diseases \cite{std_1}, socially influenced behavior \cite{social_pandemic}, and airborne diseases \cite{airborne_sample_1}. Airborne pandemics have raised the most concern due to their high infection rates and the wide population they can affect without restriction. This category of airborne pandemics includes various pathogens such as influenza, Lassa virus, COVID-19, and Nipah virus, among others \cite{airborne_alot}.

Modeling pandemic spread and control is a complex interdisciplinary endeavor that requires understanding various domains like mathematics, biology, medicine, sociology, and economics; all of which influence or are influenced by pandemics and pandemic intervention policies (PIPs) \cite{model_more_knowledge,ijerph192316023}. These models can generally be categorized into statistical models and mechanistic models. Statistical models rely on data-driven approaches without specific assumptions on dynamics, often using statistical and machine learning methods for forecasting some properties related to the pandemic \cite{stat_model_pandemic,different_approach_from_sir}. On the other hand, mechanistic models, commonly exemplified by the Susceptible-Infected-Recovered (SIR)  model, operate based on theoretical principles to explain pandemic dynamics \cite{sir_popular_4}. For instance, the SIR model assumes that the population is divided into susceptible, infected, and recovered individuals such that each infected individual infected on average \(\beta\) susceptible individuals and that infected individuals recover and become recovered at a rate \(\gamma\) \cite{first_sir}. the SIR model's simplicity often falls short in capturing real-world pandemic dynamics \cite{sir_too_simple}, leading to the development of numerous extensions and adaptations \cite{review_zika,review_covid}. These extensions are typically divided into spatial and temporal categories. Spatial extensions are represented either by norm-based or graph-based models, addressing the well-mixed population assumption used in the SIR model. Temporal extensions introduce more types of interactions and properties for individuals in the population, which play a role in the spread of the pandemic \cite{teddy_review}.

Recently, several spatio-temporal extended SIR-based models have been proposed to better capture the spread of the COVID-19 pandemic \cite{teddy_buildings,teddy_economic,spatial_1,spatial_2,spatial_3}. For instance, \cite{norm_based_1} proposed a partial differential equation (PDE) with the SEIRD (E - exposed, D - dead) model based spatio-temporal. Different diffusion parameters are also factored in for various population groups with spatial movement across a large population modeled using an inhomogeneous random walk, converging to a second-order differential operator in the limit. The authors applied this model specifically to the COVID-19 outbreak in Lombardy, Italy, focusing on the initial six months of the pandemic. \cite{graph_based} introduced a location graph-based SIR model where each graph node represents a location inhabited by a subset of the population. In their model, susceptible individuals can become infected either through contact with infectious individuals at the same node (local infection) or through infectious individuals from neighboring nodes who visit their node (global infection). The authors also factored in social connectivity, which represents the average number of individuals moving between nodes, thus facilitating global infection. In our model, we follow this trend, proposing a graph-based extended SIR model for the pandemic spread.

Moreover, multiple models also take into consideration the complex relationship between healthcare systems and pandemics \cite{sir_hosp_g_1,sir_hosp_g_2}. For example, \cite{hosp_exp_1} used a stochastic extended SIRD model with symptomatic infectious, asymptomatic infectious, and hospitalized individuals such that the model's parameters are binomial distributed rather than constant. The authors used data from COVID-19 in Texas (US) to predict the number of hospitalized individuals over time due to the pandemic. Similarly, \cite{hosp_exp_2} integrated the average quality of treatment into the SIR model to capture the pandemic spread with different levels of healthcare systems. In our model, we also include a healthcare system, extending it to handle both pandemic-related patients and soldiers with war injuries. 

\section{Model Definition}
\label{sec:model}
\begin{figure}[!ht]
    \centering
    \includegraphics[width=0.99\textwidth]{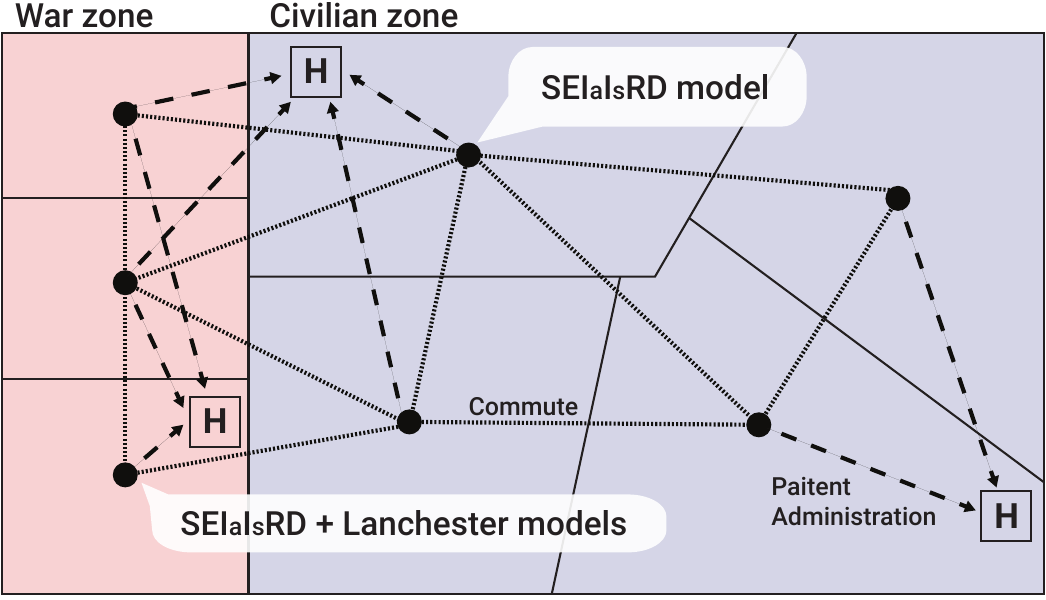}
    \caption{A schematic view of the pandemic spread with a dual healthcare system during war model. The dynamics can be divided into four interconnected components: pandemic spread (\(SEI^aI^sRD\) model), war-related death and wounded soldiers (Lanchester model), commute between locations (graph-based model), and healthcare system (functional model). }
    \label{fig:model}
\end{figure}

During times of war, countries are divided into civilian and war zones \cite{sec_3_1}, each hosting distinct yet interconnected populations of civilians and soldiers. In civilian zones, individuals engage in various activities such as commuting for work, accessing social services, or leisure purposes like tourism \cite{sec_3_2}. Similarly, war zones consist of multiple locations where soldiers are deployed and often commute between, either for operational duties or to civilian areas for rest and recuperation. The effectiveness of soldiers in fulfilling their duties in war zones is critically dependent on adequate personnel availability \cite{sec_3_3}. Injured soldiers in war zones are evacuated to healthcare centers—either military or civilian—where the proximity of the facility to the site of injury significantly impacts the quality of care and recovery outcomes \cite{sec_3_8,sec_3_7}. Compounding the challenges of war, pandemics can emerge, affecting both soldiers and civilians. Infected individuals may present as asymptomatic or symptomatic, with the latter at greater risk of death without timely medical intervention \cite{sec_3_4}. To improve survival rates, symptomatic individuals often require hospitalization, emphasizing the importance of accessible healthcare during such crises \cite{sec_3_5,sec_3_6}. Fig. \ref{fig:model} presents a schematic view of the pandemic spread with a dual healthcare system during a war model.

To capture the epidemiological dynamics during warfare involving civilian and military populations within a dual-use healthcare system, we propose a spatio-temporal framework that integrates an extended SIR model for epidemiological dynamics and the Lanchester model for warfare dynamics. For simplicity, we divide the proposed model into four components: spatial movement, pandemic spread, hospitalization, and warfare. Based on these components, we assemble a single mathematical framework. 

Formally, we define the spatio-temporal model that captures these dynamics as follows. A model, \(M\), is defined by the tuple \(M := (G, P, H, W)\) such that \(G\) is the graph of locations and possible commute between them, \(P\) is the pandemic spread dynamics, \(H\) is the association of healthcare centers with locations and their treatment dynamics, and \(W\) is the war-related dynamics. Below, we define each component independently and then combine them into a single framework. 

\subsection{Spatial graph representation}
Let us consider a graph \(G := (V_w, V_c, E)\) where \(V_w\) is the set of war zone locations, \(V_c\) is the set of civilian locations, and \(E\) is the set of possible commutes between all locations, such that \(E \subseteq (V_w \cup V_c) \times (V_w \cup V_c) \times \mathbb{R}^2 \). Each location contains two sub-populations - a civilian and a military. In terms of the pandemic spread, the populations are well-mixed\footnote{The well-mixed property assumes that at any point in time, any two individuals in the population have the same probability to interact.} and heterogeneous. For the commute dynamics, each edge \(e := (i, j. c_w, c_c) \in E\) between nodes \(i\) and \(j\), has two weights \(c_w\) and \(c_c\) indicating the average commute between the two locations for military and civilian sub-populations, respectively. In practice, \(c_w\) and \(c_c\) are time-dependent, however, for simplicity, we use the average value over time. 

\subsection{Pandemic spread dynamics}

\begin{figure}[!ht]
    \centering
    \includegraphics[width=0.99\textwidth]{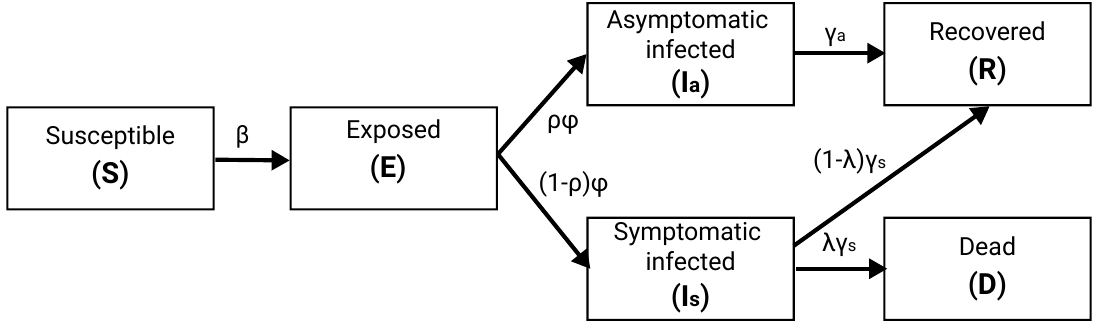}
    \caption{A schematic view of the pandemic spread model presenting the epidemiological states and the transformations between them. }
    \label{fig:sir_model}
\end{figure}

We adopted the \(SEI_aI_sRD\) model \cite{pandemic_management_teddy,seird_1,seird_2}. Namely, the model considers a constant population with a fixed number of individuals \(N\). For simplicity, and given the short time horizon of interest, we abstract from population growth. Each individual belongs to one of the six groups: susceptible \((S\)),  exposed \(E\), asymptomatic infected \((I^a\)), symptomatic infected \((I^s\)), recovered \((R)\), and dead \((D)\) such that \(N = S + E + I^a + I^s + R + D\). Individuals in the first group have no immunity and are susceptible to infection. When an individual in the susceptible group (\(S\)) is exposed to the pathogen, the individual is transferred to the exposed state (\(E\)) at rate  \(\beta\). The individual stays in the exposed state for \(\psi\) time steps in time and then transforms to either the asymptomatic infected group (\(I^a\)) or symptomatic infected group (\(I^s\)) with rate \(\rho\) and \(1-\rho\), respectively. The individual stays in the symptomatic infected group on average \(\gamma_s\) time steps, after which the individual is transferred to the recovered group (\(R\)) or the dead group \((D)\) with rate \(\lambda\) and \(1-\lambda\), respectively. All asymptomatic infected individuals stay in the infected \(I^a\) group on average \(\gamma_a\) time steps, after which the individual is transferred to the recovered group \((R)\). The recovered are again healthy, no longer contagious, and immune from future infection. The epidemiological dynamics are formally described below using a system of ordinary differential equations:
\begin{equation}
    \begin{array}{l}
         \frac{dS(t)}{dt} = -\beta S(t) \big ( I_s(t) +  I_a(t) \big ),   \\\\
         \frac{dE(t)}{dt} = \beta S(t) \big ( I_s(t) +  I_a(t) \big ) - \psi E(t),  \\ \\
         \frac{dI_s(t)}{dt} =  \rho \psi E(t) - \gamma_s I_s(t), \\\\
          \frac{dI_a(t)}{dt} =  (1 - \rho) \psi E(t) - \gamma_a I_a(t), \\\\
           \frac{dR(t)}{dt} =  \gamma_a I_a(t) + \lambda \gamma_s I_s(t), \\\\
           \frac{dD(t)}{dt} = (1 - \lambda) \gamma_s I_s(t). \\ \\
    \end{array}
    \label{eq:epi_model}
\end{equation}
Fig. \ref{fig:sir_model} presents a schematic view of the pandemic spread models, divided into its epidemiological states. 

\subsection{Hospitalization dynamics}
\begin{figure}[!ht]
    \centering
    \includegraphics[width=0.5\textwidth]{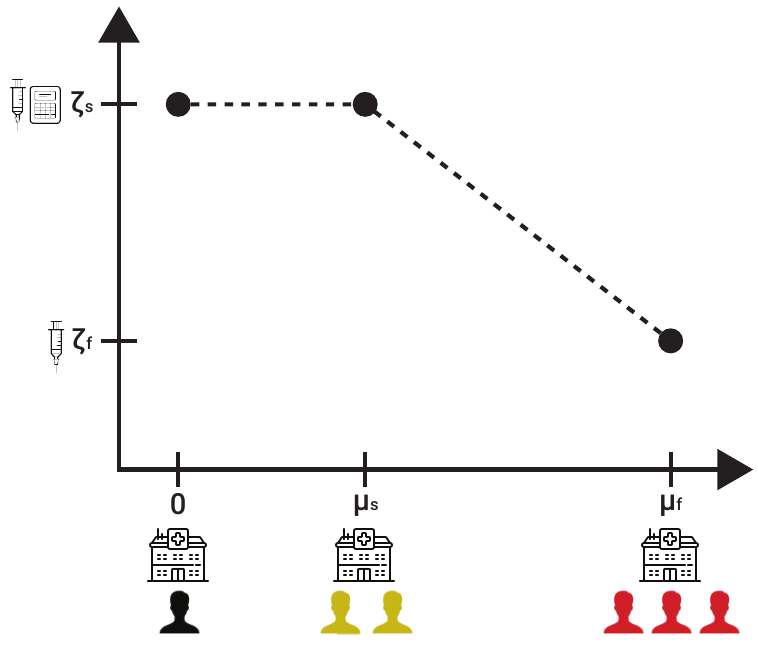}
    \caption{A schematic view of the hospitalization performance as a function of the number of administrated individuals. }
    \label{fig:hospatalization_profile}
\end{figure}

The healthcare system is based on a set of healthcare service provider locations. Each healthcare service provider's location is able to treat a number of individuals (\(\mu_s\)) while providing the best clinical treatment possible which reduces the probability of dying due to the pandemic or war injuries (\(\lambda, \epsilon\)) in a rate \(\zeta_s\). However, once the number of individuals administrated to the healthcare center extends \(\mu_s\), the treatment performance is linearly decreasing until \(\mu_f\) individuals are administrated to the healthcare center where the treatment performance is reaching \(\zeta_f\). In other words, \(\Psi: \mathbb{N} \rightarrow [0, 1]\), indicates the reduction in the probability of dying due to the pandemic (\(\lambda\)) as a function of the number of administrated individuals:
\begin{equation}
    \Psi(x) := \begin{cases}
        \zeta_s \; \text{ if } x < \mu_s \\ 
        \zeta_s - \frac{\zeta_f = \zeta_s}{\mu_f - \mu_s}(x - \zeta_s) \; \text{ if } \mu_s \leq x \leq \mu_f \\  
        0 \; \text{ if } x > \mu_f \\ 
    \end{cases}
    \label{eq:hospitalization}
\end{equation}
Fig. \ref{fig:hospatalization_profile} presents a schematic view of the hospitalization performance as a function of the number of administrated individuals. 

\subsection{War dynamics}
To capture the war dynamics, we adopted the Lanchester model \cite{war_model} and added wounded soldiers dynamics. Formally, we assume that there are only two combating sides and that the two sides comprise a homogeneous force. Let us consider \(A(t)\) and \(B(t)\) to be the sizes of the two armies at time \(t\). Each side has an average performance of reducing the size of the other army denoted by \(\epsilon_a\) and \(\epsilon_b\), respectively. From the population that was removed from each army at some point in time \(t\), a portion is only wounded with injury rates of \(\theta_a\) and \(\theta_b\), respectively. Wounded soldiers do not participate in the war and die out at a rate \(d\) if not treated. Formally, this dynamic takes the form:
\begin{equation}
    \begin{array}{l}
         \frac{dA(t)}{dt} = -\epsilon_b B(t), \\ \\
         \frac{dB(t)}{dt} = -\epsilon_a A(t), \\ \\
         \frac{dW^a(t)}{dt} = \theta_b \epsilon_b B(t) - dW^a(t), \\ \\
         \frac{dW^b(t)}{dt} = \theta_a \epsilon_a A(t) - dW^b(t). \\ \\
    \end{array}
    \label{eq:war_dynamics}
\end{equation}

\subsection{Assembling the components into a single framework}
To assemble the four components into a single framework, one has to outline the relationship between the components. First, the graph-based spatial component introduces a spatial separation to local interaction - either for the war or pandemic dynamics as well as a commute between the locations, it enforces the later dynamics to occur on each node of the graph and includes movement of the population. Next, the pandemic's spread should consider both soldiers and civilians to fit the war dynamics and the commute dynamics that occur in parallel. The pandemic and war occurring for both armies in each location are combined into a single process. Finally, both symptomatic infected individuals and wounded soldiers are administrated to healthcare service providers according to some policy. 

To this end, the framework is presented at the node level such that the dynamics in all the nodes occur in parallel over time. Since there are both civilian and war-related nodes, we describe the dynamics in each separately. Starting with the civilian node, the dynamics take the following form: 

 \begin{equation}
    \begin{array}{l}
    \frac{dS^c(t)}{dt} = -\beta S^c(t) \big ( I^{c}_s(t) + I^{c}_a(t) + I^{m}_s(t) + I^{m}_a(t) \big ) + Commute(S^c) ,   \\\\
    
    \frac{dE^c(t)}{dt} = \beta S^c(t) \big ( I^{c}_s(t) + I^{c}_a(t) + I^{m}_s(t) + I^{m}_a(t) \big ) - \psi E^c(t) + Commute(E^c),  \\ \\
    
    \frac{dI^{c}_s(t)}{dt} = \rho \psi E^c(t) - \gamma_s I^{c}_s(t) + Commute(I_s^c) - hosp(H^{c}(t), I^{c}_s(t)), \\\\
    
    \frac{dI^{c}_a(t)}{dt} = (1 - \rho) \psi E^c(t) - \gamma_a I^{c}_a(t)+ Commute(I_a^c), \\\\
    
    \frac{dR^c(t)}{dt} = \gamma_a I^{c}_a(t) + \lambda \gamma_s I^{c}_s(t) + (\lambda + \Psi(H^c(t) + H^m(t)))H^c(t) + Commute(R^c), \\\\
    
    \frac{dD^c(t)}{dt} = (1 - \lambda) \gamma_s I^{c}_s(t) + (1 - \lambda - \Psi(H^c(t) + H^m(t)))H^c(t), \\ \\
    
    \frac{dH^c(t)}{dt} = hosp(H^{c}(t), I^{c}_s(t)) - \gamma_s H^c(t), \\\\
    
    \frac{dS^m(t)}{dt} = -\beta S^m(t) \big ( I^{c}_s(t) + I^{c}_a(t) + I^{m}_s(t) + I^{m}_a(t) \big ) + Commute(S^m),   \\\\
    
    \frac{dE^m(t)}{dt} = \beta S^m(t) \big ( I^{c}_s(t) + I^{c}_a(t) + I^{m}_s(t) + I^{m}_a(t) \big ) - \psi E^m(t) + Commute(E^m),  \\ \\
    
    \frac{dI^{m}_s(t)}{dt} = \rho \psi E^m(t) - \gamma_s I^{m}_s(t) + Commute(I_s^m) - hosp(H^{m}(t), I^{m}_s(t)), \\\\
    
    \frac{dI^{m}_a(t)}{dt} = (1 - \rho) \psi E^m(t) - \gamma_a I^{m}_a(t) + Commute(I_a^m), \\\\
    
    \frac{dR^m(t)}{dt} = \gamma_a I^{m}_a(t) + \lambda \gamma_s I^{m}_s(t) + (\lambda + \Psi(H^c(t) + H^m(t)))H^m(t)  + Commute(R^m), \\\\
    
    \frac{dD^m(t)}{dt} = (1 - \lambda) \gamma_s I^{m}_s(t) + (1 - \lambda - \Psi(H^c(t) + H^m(t)))H^m(t), \\\\
    
    \frac{dH^m(t)}{dt} = hosp(H^{m}(t), I^{m}_s(t)) - \gamma_s H^m(t), \\\\
\end{array}
     \label{eq:model_civial_node}
 \end{equation}
where \(n := |V_w| + |V_c|\) is the total number of nodes, \(m_{i,j}\) is the average commute between nodes \(i\) and \(j\), \(Commute(X) := \sum_{j=1, j \neq k}^n \big ( m_{j,k} X_j(t) \big ) - \sum_{j=1, j \neq k}^n \big ( m_{k,j} X(t) \big ) \), and \(hosp(x, y)\) is the patient administration policy for symptomatically infected individuals.

For the war nodes, the dynamics occur for two army groups as it is assumed no citizens of either side of the war are present in such locations. Thus, the dynamics take the form:

\begin{equation}
    \begin{array}{l}
        \frac{dS^a(t)}{dt} = -\beta^a S^a(t) \big ( I^{a}_s(t) + I^{a}_a(t) \big ) - \beta^{ab} \big (I^{b}_s(t) + I^{b}_a(t) \big ) - \epsilon_b \frac{S^a(t)\Lambda_b(t)}{\Lambda_a(t)} + Commute(S^a) ,   \\\\
        
        \frac{dE^a(t)}{dt} = \beta^a S^a(t) \big ( I^{a}_s(t) + I^{a}_a(t) \big ) + \beta^{ab} \big (I^{b}_s(t) + I^{b}_a(t) \big ) - \psi E^a(t) - \epsilon_b \frac{E^a(t)\Lambda_b(t)}{\Lambda_a(t)}  + Commute(E^a),  \\ \\
        
        \frac{dI^{a}_s(t)}{dt} = \rho \psi E^a(t) - \gamma_s I^{a}_s(t) + Commute(I_s^a) - hosp(H^{a}(t), I^{a}_s(t), W^{a}(t)), \\\\
        
        \frac{dI^{a}_a(t)}{dt} = (1 - \rho) \psi E^a(t) - \gamma_a I^{a}_a(t) - \epsilon_b \frac{I^a_a(t)\Lambda_b(t)}{\Lambda_a(t)} + Commute(I_a^a), \\\\
        
        \frac{dR^a(t)}{dt} = \gamma_a I^{a}_a(t) + \lambda \gamma_s I^{a}_s(t) - \epsilon_b \frac{R^a(t)\Lambda_b(t)}{\Lambda_a(t)} + (\lambda + \Psi(H^a(t)))\gamma_s H^a(t) + \Psi(H^a(t))\gamma_w H^a(t) + Commute(R^a), \\\\
         
         \frac{dW^a(t)}{dt} = \theta_b \epsilon_b \Lambda_b(t) - dW^a(t) - hosp(H^{a}(t), I^{a}_s(t), W^{a}(t)), \\ \\
        
        \frac{dD^a(t)}{dt} = (1 - \lambda) \gamma_s I^{a}_s(t) - (1 - \lambda - \Psi(H^a(t))\gamma_s H^a(t), \\ \\

        \frac{dH^a(t)}{dt} = hosp(H^{a}(t), I^{a}_s(t), W^{a}(t)) - \gamma_s H^a(t) - \gamma_w H^a(t), \\\\
        
        \frac{dS^b(t)}{dt} = -\beta^b S^b(t) \big ( I^b_s(t) + I^b_a(t) \big ) - \beta^{ab} \big (I^{a}_s(t) + I^{a}_a(t) \big ) - \epsilon_a \frac{S^b(t)\Lambda_a(t)}{\Lambda_b(t)} + Commute(S^b) ,   \\\\
        
        \frac{dE^b(t)}{dt} = \beta^b S^b(t) \big ( I^b_s(t) + I^b_a(t) \big ) + \beta^{ab} \big (I^{b}_s(t) + I^{b}_a(t) \big ) - \psi E^b(t) - \epsilon_a \frac{E^b(t)\Lambda_a(t)}{\Lambda_b(t)}  + Commute(E^b),  \\ \\
        
        \frac{dI^b_s(t)}{dt} = \rho \psi E^b(t) - \gamma_s I^b_s(t) + Commute(I_s^b), \\\\
        
        \frac{dI^b_a(t)}{dt} = (1 - \rho) \psi E^b(t) - \gamma_a I^b_a(t) - \epsilon_a \frac{I^b_a(t)\Lambda_a(t)}{\Lambda_b(t)} + Commute(I_a^b), \\\\
        
        \frac{dR^b(t)}{dt} = \gamma_a I^b_a(t) + \lambda \gamma_s I^b_s(t) - \epsilon_a \frac{R^b(t)\Lambda_a(t)}{\Lambda_b(t)}+ (\lambda + \Psi(H^b(t)))\gamma_s H^b(t) + \Psi(H^b(t))\gamma_w H^b(t)   + Commute(R^b), \\\\
         
         \frac{dW^b(t)}{dt} = \theta_a \epsilon_a \Lambda_a(t) - dW^b(t) - hosp(H^{b}(t), I^{b}_s(t), W^{b}(t)), , \\ \\
        
        \frac{dD^b(t)}{dt} = (1 - \lambda) \gamma_s I^b_s(t)- (1 - \lambda - \Psi(H^b(t))\gamma_s H^b(t), \\ \\
    
        \frac{dH^b(t)}{dt} = hosp(H^{b}(t), I^{b}_s(t), W^{b}(t)) - \gamma_s H^b(t) - \gamma_w H^b(t), \\\\
        
    \end{array}
    \label{eq:model_war_node}
\end{equation}
where \(\beta^a\), \(\beta^b\), and \(\beta^{ab}\) are the infection rates of the first army, second army, and between armies. We denote \(\Lambda_a(t) = S^a(t) + E^a(t) + I^a_a(t) + R^a(t)\) and \(\Lambda_b(t) = S^b(t) + E^b(t) + I^b_a(t) + R^b(t)\).
 
\section{Computer Simulation}
\label{sec:simulator}
In this section, we present an agent-based simulation (ABS) approach to solve the proposed model (Eq. (\ref{eq:model_civial_node}-\ref{eq:model_war_node})) followed by a reinforcement learning-based model to derive a (near-)optimal patient administration policy.

\subsection{System solver using agent-based simulation}
We implemented the proposed model (Eq. (\ref{eq:model_civial_node}-\ref{eq:model_war_node})) using the ABS approach following the scheme proposed by \cite{teddy_review}. Formally, let us assume a population of agents allocated to a graph of location (\(G\)) in some pre-defined distribution. These agents move and interact in discrete finite time steps \(t \in [1, \dots, T]\), where \(T<\infty\). In order to use the ABS approach, one has to define the agents in the dynamics as well as their three types of interactions: agent-agent, agent-environment, and spontaneous (i.e., depends only on the agent's state and time) \cite{agent_based_exp_1}. To this end, for our model, each agent is represented by a timed finite state machine \cite{fsm} where its epidemiological-clinical status (\(S, E, I^c, I^a, R, D, W, H\)), current location (\(v \in V_m \cup V_c\)), and sociological status (solder or civilian). The simulation starts by generating the location graph and allocating the agents inside it with their walking dynamics, as indicated by the average commute rate between each two locations, and epidemiological state. Next, iteratively, until a stop condition is met or a pre-defined number of steps in time is reached, the simulation solves the dynamics of each node, in a random order. First, the portion of the population in each node moves to other nodes according to the average commute rate between locations. Afterward, the epidemiological and war-related dynamics take place (Eq. (\ref{eq:model_civial_node}-\ref{eq:model_war_node})). Finally, individuals are allocated to healthcare centers according to some patient administration policies. Notably, healthcare centers and locations can be associated in many-to-many manner. 

\subsection{Patient administration policy optimization}
In this section, we outline a deep reinforcement learning (DRL) model to determine the optimal patient administration policy aimed at minimizing overall mortality, considering deaths from both war-related causes and the pandemic. The RL method leverages the agent-based simulation (ABS) model described earlier to iteratively improve decision-making regarding the allocation of patients to healthcare centers.

The goal of the DRL method is to identify a patient administration policy \(\pi\) that minimizes the cumulative death count over the simulation period \(t \in [0, T]\). The administration policy \((\pi)\) determines the actions (patient administration) based on the current state of the system, which includes the epidemiological-clinical status of individuals, their locations, and the capacity and status of healthcare centers. Formally, the RL model's state space at time \(t\) (\(s_t\)) is a comprehensive representation of the current situation, including, the epidemiological status of individuals (\(S, E, I^s, I^a, R, D, W, H\)), their current location (\(v \in V_m \cup V_c\)), sociological status, and current capacity and occupancy of healthcare centers, as well as the state of the other army in the war zones. The action space, \(A\), is composed of the set of patient administrations from all nodes in the graph to all healthcare centers and represented by a tensor of size \(|V_m \cup V_c| \times |C| \times 2\) such that \(|C|\) indicates the number of healthcare centers where each value in the tensor is a two-dimensional vector indicating the number of civilians and soldiers allocated to each of the healthcare centers. The reward function of the RL model for time \(t\) is
\begin{equation}
    R(s_t, a_t) = - \sum_{v \in V_m \cup V_c} \big ( D_v(t) + W_v(t) \big ).
    \label{eq:reward}
\end{equation}
We employ a deep Q-network (DQN) algorithm \cite{q_learning}, a model-free, online, and off-policy RL method, to optimize the patient administration policy. A DQN algorithm is a value-based RL method that trains a critic to estimate the expected discounted cumulative long-term reward when following the optimal administration policy. For the neural network architecture, we used three hidden layers including an LSTM with 64 neurons and two fully connected layer of sizes 64 and 32 with ReLu activations. The output layer's size corresponds to the number of healthcare centers (\(|C|\)). Notably, during the training phase, actions involving nodes that cannot allocate civilians and/or soldiers to specific hospitals, or to hospitals that are already at full capacity, were masked by assigning a large negative value to their corresponding entries in the action space (\(-10^9\)). 

The RL algorithm is implemented within the ABS framework, enabling the simulation of various patient administration strategies and their impacts on overall mortality. Through iterative learning, the administration policy \(\pi\) converges towards an optimal strategy that effectively allocates patients to minimize deaths from both war-related causes and the pandemic. Formally, in order to allow the RL model to learn a representative policy, we repeat each ABS configuration multiple times and then alter the ABS configuration. Table \ref{table:appendix_parameters_rl} presents the parameter values of the DRL model. The parameter values are chosen using a manual trial-and-error to achieve a good administration policy's performance. 

\begin{table}[!ht]
    \centering
    \begin{tabular}{llc}
        \hline \hline
        \textbf{Hyperparameter} & \textbf{Description} & \textbf{Default value} \\
        \hline \hline
        Learning Rate & Step size for updating the Q-values & 0.001 \\
        
        Discount Factor & Factor to discount future rewards & 0.98 \\
        
        Exploration Rate & Probability of choosing a random action & 0.15 \\
        
        Min. Exploration Rate & Minimum value of the exploration rate & 0.02 \\
        
        Exploration Decay Rate & Rate at which the exploration rate decays & 0.985 \\
        
        Replay Buffer Size & Number of experiences stored in the replay buffer & 104 \\
        
        Batch Size & Number of experiences sampled from the replay buffer & 8 \\
        
        Number of Episodes & Total number of episodes for training & 100 \\
        
        Maximum Steps per Episode & Maximum number of steps per episode & 2160  \\ \hline \hline
    \end{tabular}
    \caption{Hyperparameters for training the Q-learning model.}
    \label{table:appendix_parameters_rl}
\end{table}

\section{Evaluation}
\label{sec:analysis_numerical}
To evaluate the proposed model, we first established realistic, while synthetic, scenarios. The sociological and movement parameters such as the population size and number of nodes in the graphs (which represent central cities) aim to capture a large-scale region up to the entire western country \cite{weber2019effect}. For the pandemic's spread, we focused on the COVID-19 pathogen due to the large and accurate recorded data associated with this pathogen \cite{betthaus2023systematic}. The warfare and hospitalization parameters ranged greatly in the literature. As such, we adopted a range without anomalies. Table \ref{table:parameters_model} summarizes the parameters and their values as these are used by the proposed model. 

\begin{table}[h]
    \centering
    \begin{tabular}{p{0.06\textwidth}p{0.60\textwidth}p{0.16\textwidth}p{0.1\textwidth}}
        \hline \hline
        \textbf{Symbol} & \textbf{Description} & \textbf{Value range} & \textbf{Source} \\
        \hline \hline
        \(N\) &  Initial population size & \(10^6 - 10^8\) & Assumed \\
        \(\) &  Portion of the population participating in the army & \(0.01 - 0.1\) & Assumed \\
        \(\) & Number of hospitals & 1-6 & Assumed \\
        \(|V_m|\) & Number of war-zone nodes of the graph & 1-20 & Assumed \\
        
        \(|V_c|\) &  Number of civilian nodes of the graph & 10-100 & Assumed \\
        
        \(T\) & Total number of simulation steps & 200 & Assumed \\
        
        \(\Delta t\) & Time steps in each simulation iteration & 1 hour & Assumed \\
        
        \(c_w, c_c\) & Average commute rates between locations & \(1 \cdot 10^{-5}-5 \cdot 10^{-4}\) & \cite{move_values} \\
        
        \(\beta\) & Average infection rate & \(0.0014 - 0.0072\) & \cite{pandemic_management_teddy} \\

        \(\rho\) & Symptomatic rate  & \(0.01-0.1\) & \cite{pandemic_management_teddy} \\

        \(\lambda\) & Death rate & \(0.0014 - 0.0072\) & \cite{pandemic_management_teddy} \\

        \(\psi\)  & Average incubation time in days spent in the exposed state & \(4-7\) & \cite{pandemic_management_teddy} \\
        
        \(\gamma_s\)  & Average time in days spent in the symptomatically infected state & \(10-18\) & \cite{pandemic_management_teddy} \\
        
        \(\gamma_a\)  & Average time in days spent in the asymptomatic infected state & \(7-10\) & \cite{pandemic_management_teddy} \\
        
        \(\mu_s\) &  Maximum number of patients hospitals can treat without effect on clinical performance & \(1000 - 10000\) & Assumed \\
        
        \(\mu_f\) &  Maximum number of patients hospitals can treat over \(\mu_s\) & \(125\% - 150\%\) & Assumed \\
        
        \(\zeta_s\) & The baseline rate of recovery from pandemic or war injuries & \(0.95-0.99\) & \cite{clinical_3} \\
        
        \(\zeta_f\) & The baseline rate of recovery from pandemic or war injuries when the hospital is full & \(0.8-0.95\) & \cite{clinical_3} \\
        
        \(\epsilon_a, \epsilon_b\) & Military death rate due to war & \(1 \cdot 10^{-5} - 5 \cdot 10^{-5}\) & \cite{war_values} \\
        
        \(\theta_a, \theta_b\) & Military injury rate due to war & \(1 \cdot 10^{-5} - 5 \cdot 10^{-5}\) & \cite{war_values} \\

        \(d\) & Death rate of wounded soldiers that were not treated & \(4.07    \cdot 10^{-5}\) & \cite{lethality_values} \\
        \hline \hline
    \end{tabular}
    \caption{The model's parameters' description, value ranges, and sources. }
    \label{table:parameters_model}
\end{table}

Initially, we train the proposed DRL model to learn a (near-)optimal policy for patient administration. Fig. 5 shows the relative performance of the DRL model, with aims to decrease the total number of deaths from both the pandemic and war (see Eq. (\ref{eq:reward})), with respect to the normalized random allocation (baseline) that is computed for \(n=100\) random simulations. Initially, the DRL's performance is comparable to the baseline one, as it is also initialized with random logic. Initially, for the first 80 simulations, the DRL performance was worse than the baseline, up to 35\% worse. However, after this point, a relatively smooth exponential improvement in the DRL's performance occurs, conversing to around 30\% of the baseline performance (i.e., 70\% improvement) after 800 simulations. Notably, around the 300-350 and 625-675 simulations, a local divergence takes place which can be associated with a local optimum of the optimization process converge and requires more exploration to overcome it and re-converge to more global optima \cite{zhang2020global}. 

\begin{figure}
    \centering
    \includegraphics[width=0.99\linewidth]{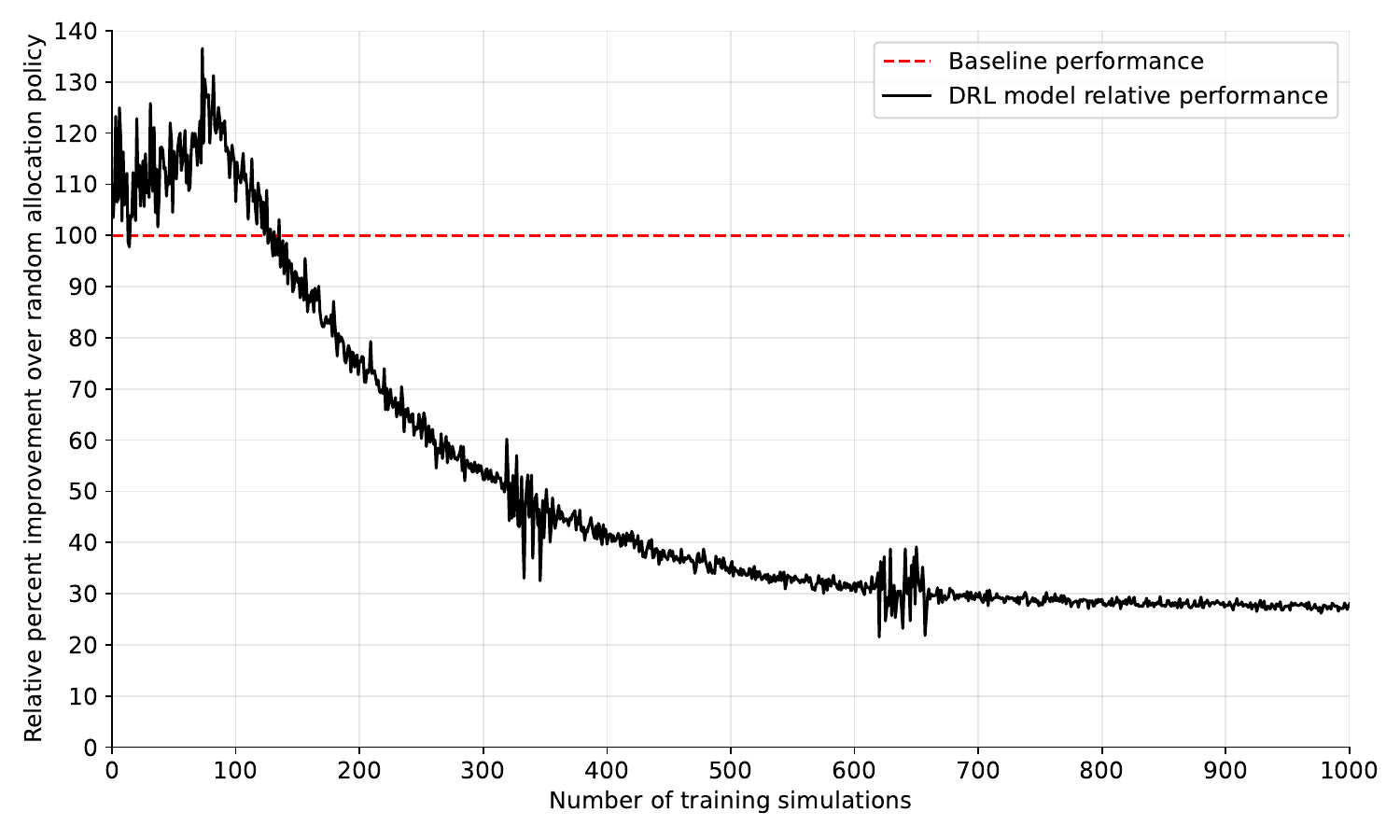}
    \caption{The DRL's administration policy's average performance normalized to the naive administration policy for \(n=100\) simulations as a function of the number of training simulations.}
    \label{fig:drl_training_graph}
\end{figure}

\begin{figure}
    \centering
    \includegraphics[width=0.99\linewidth]{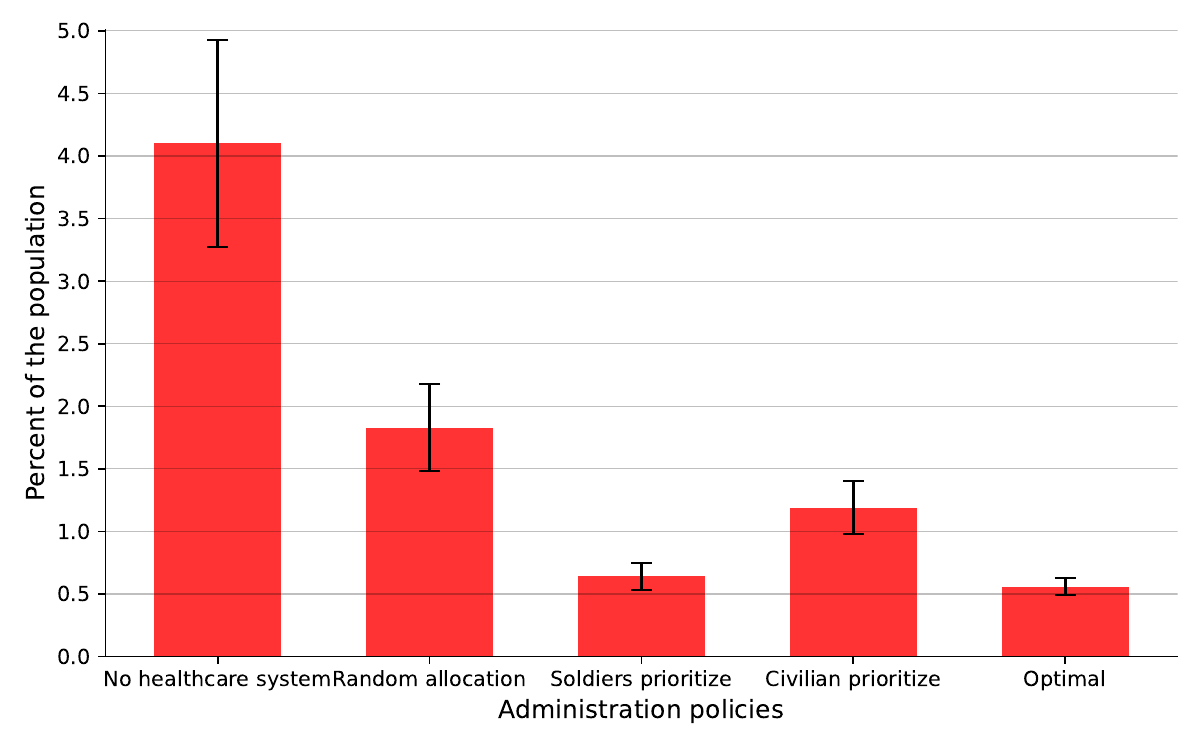}
    \caption{Overall death as a function of different administration policies. The results are shown as the mean \(\pm\) standard deviation of \(n=100\) simulations.}
    \label{fig:drl_types_differances}
\end{figure}

Next, to evaluate the performance of the DRL model we compared it to two baseline policies and two single-usage healthcare polices. Fig. \ref{fig:drl_types_differances} shows the overall death as the mean \(\pm\) standard deviation of \(n_{test}=100\) simulations following a \(n_{train} = 1000\) simulation to train the DRL model for each case, divided into four patient administration policies - no patient administration, random allocation up to full capacity, optimized based on Eq. (\ref{eq:reward}), DRL model optimized to reduce the death of soldiers alone, and DRL model optimized to reduce the death of civilians alone. The optimal configuration is found to be statistically significantly better based on ANOVA (Analysis of Variance) with Tukey post-hoc test \cite{anova} with \(p < 0.05\). Unsuspectingly, the case where no healthcare system is significantly worse than the other cases with slightly over 4\% death of the population. The naive random allocation results in less than half of the percent of death while also the standard deviation is smaller, as indicated by the error bars. Moreover, comparing the single-usage healthcare cases, in the form of the soldiers and civilians alone cases, the latter is statistically significantly worse with \(p < 0.01\) based on a Mann–Whitney U test \cite{mcknight2010mann}.

\begin{figure}
    \centering
    \includegraphics[width=0.99\linewidth]{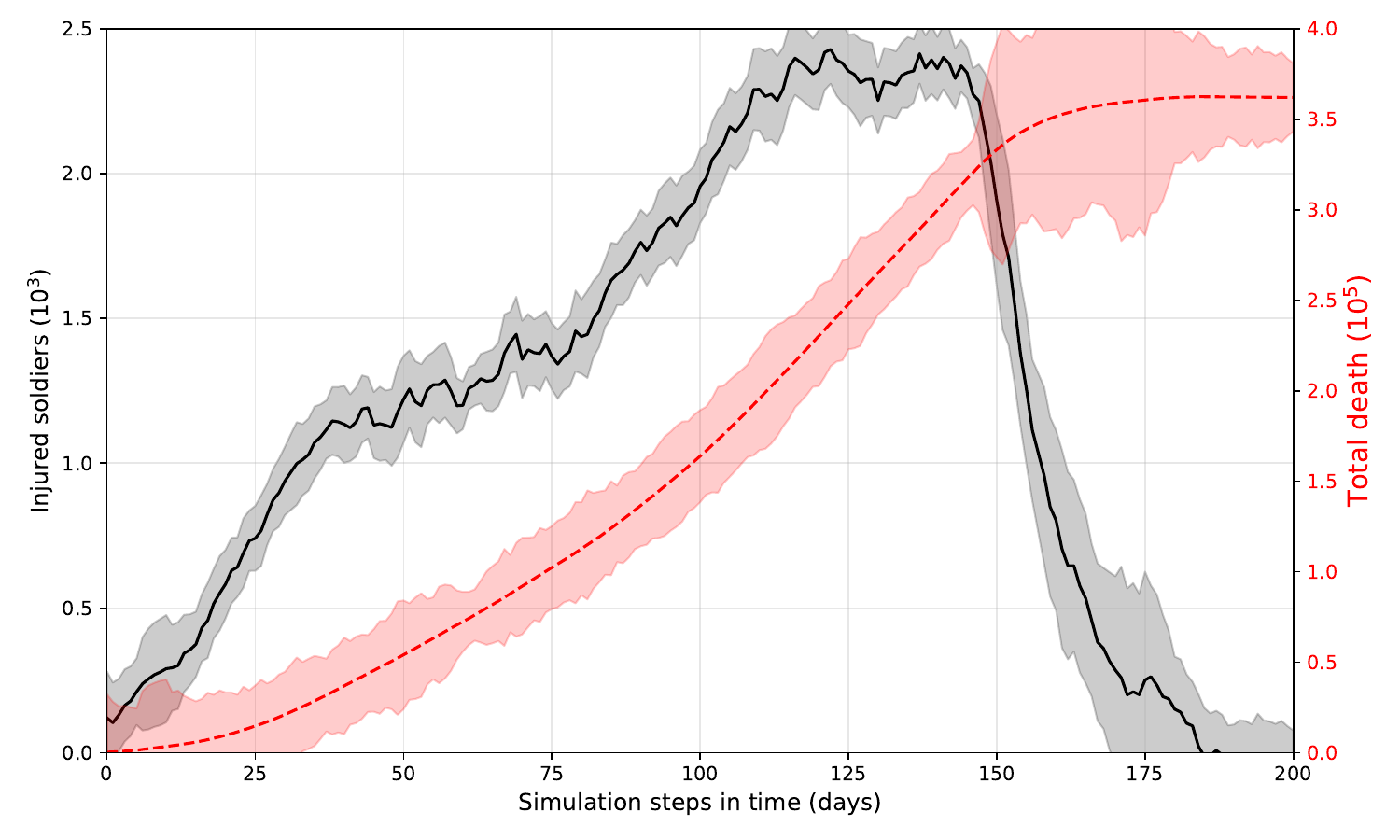}
    \caption{The wounded soldiers and overall deaths (both soldiers and civilians) over time using the optimal DRL administration policy. The results are shown as the mean \(\pm\) standard deviation of \(n=100\) simulations assuming a fixed population size of \(N = 5 \cdot 10^6\) with 10\% of the population are in the army..}
    \label{fig:first_graph}
\end{figure}

Fig. \ref{fig:drl_types_differances} revealed that providing healthcare services to soldiers, on average, results in overall less death in a population. Nevertheless, taking civilians into account provides a statistically significant improvement. Thus, we further explore the death and injured dynamics of soldiers over time. Fig. \ref{fig:first_graph} shows the wounded soldiers and overall deaths (both soldiers and civilians) over time under the optimal DRL administration policy. The results are presented as the mean \(\pm\) standard deviation of \(n=100\) simulations, assuming a fixed population size of \(N = 5 \cdot 10^6\) with 10\% of the population comprising army personnel. For the injured soldiers, an initial linear increase is observed up to the 35th simulation step, aligning with the expectations of the Lanchester model (see Eq. (\ref{eq:war_dynamics})). Afterward, a more chaotic pattern emerges, characterized by a slower rate of increase, which can be attributed to treated soldiers returning from the healthcare system. This behavior persists until approximately the 75th simulation step. These dynamics are followed by another increase up to the 110th simulation step, after which the values stabilize for approximately 35 simulation steps. Subsequently, a sharp decline is observed, reaching zero around the 185th simulation step. This decline can be explained by the end of the war as the other army is eliminated.In a complementary manner, the total deaths exhibit a monotonically increasing trend with a sigmoid-like pattern, reflecting the combined fatalities of injured soldiers who succumbed to war-related injuries and both soldiers and civilians who died due to the pandemic. The impact of injured soldiers on total deaths is twofold: their presence drives a faster-than-linear increase in deaths, delayed relative to their numbers. This strain on the healthcare system, occupied with treating injured soldiers, limits its capacity to manage infected individuals, exacerbating pandemic spread and mortality.

\begin{figure}
    \centering
    \includegraphics[width=0.99\linewidth]{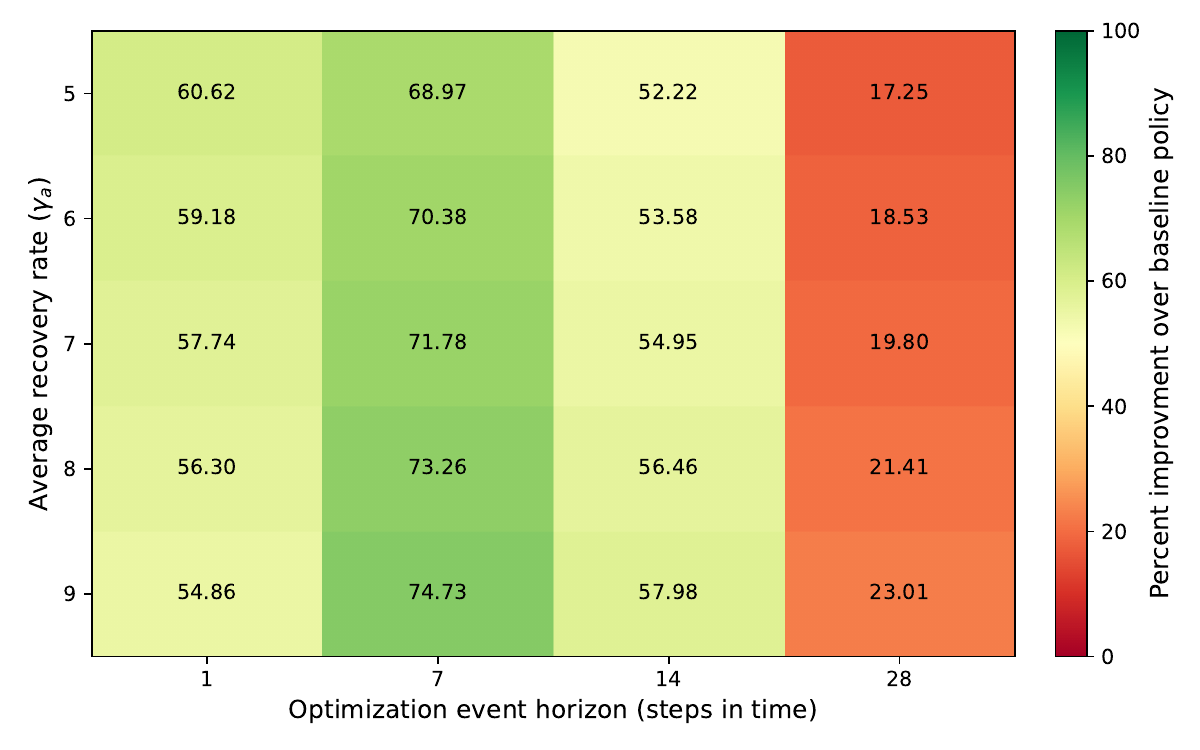}
    \caption{The mean percent improvement over the random administration (baseline) policy for \(n_{test} = 100\) simulations followed \(n_{train} = 1000\) simulations}
    \label{fig:drl_horizon}
\end{figure}

Fig. \ref{fig:first_graph} reveals non-linear and phase-dependent dynamics which further increase the uncertainty of the dynamics. Since the model assumes know-in-advance war results, in the form of the Lanchester model, the uncertainty can be associated with the pandemic spread which \say{breaks} the symmetry in the Lanchester dynamics at the node level. Thus, the event optimization horizon for which the DRL model is trained should result in different performances. To this end, Fig. \ref{fig:drl_horizon} shows the mean percent improvement over the random administration (baseline) policy for \(n_{test} = 100\) simulations followed \(n_{train} = 1000\) simulations. Notably, the model's performance improves when the event optimization horizon aligns with the average recovery rate. Interestingly, this trend holds even for a 28-day event optimization horizon, suggesting that other dynamics, such as warfare, have a secondary effect on administration policy optimization due to their greater predictability. 

\section{Conclusion}
\label{sec:discussion}
In this study, we have developed a novel spatio-temporal model to address the complex dynamics of dual-use healthcare administration policy during a pandemic in wartime conditions. By extending the SIR model to include the effects of both civilian and military populations, we offer a comprehensive framework that captures the intricate interactions between pandemic spread, war dynamics, and healthcare system performance. Our model incorporates four key components: spatial movement, pandemic spread, hospitalization dynamics, and warfare dynamics. By integrating these components, we offer a comprehensive representation of the interplay between pandemics and wars and their mutual impacts.

As an intervention strategy to reduce overall population deaths, we propose a DRL-based model that optimizes healthcare administration policy using \textit{in silico} data generated by the proposed agent-based simulation framework. The inclusion of a dual healthcare system allows for the differentiation between civilian and military medical services, highlighting the unique challenges and administrative policies in both contexts.

As shown in Fig. 5, the DRL model successfully learns an optimal administration policy, reducing the death toll by 70\% compared to a naive policy in which patients are randomly allocated, provided hospitals are not at full capacity. Moreover, the evaluation of the DRL model compared to other baseline and single-usage healthcare policies, as shown in Fig. \ref{fig:drl_types_differances}, highlights its effectiveness. The DRL policy optimized for overall population health demonstrated statistically significant improvements over other strategies. Specifically, the allocation strategy focused solely on soldiers led to fewer deaths than the one focused solely on civilians, but the combined optimization yielded the best overall results. These findings underline the importance of accounting for both military and civilian health in designing administration policies, as neglecting one population segment leads to suboptimal outcomes.

Further insights into the dynamics of death and injury under the optimal DRL administration policy are provided in Fig. \ref{fig:first_graph}. The time-dependent behavior of wounded soldiers and overall deaths reveal non-linear and phase-dependent dynamics. For example, the sharp decline in wounded soldiers around the 185th simulation step coincides with the end of the war, whereas the sigmoid-like increase in total deaths reflects the dual impact of injuries from war and deaths from the pandemic. The delay in responding to pandemic needs due to the saturation of the healthcare system by injured soldiers highlights a critical trade-off between war-related injuries and pandemic management. To this end, Fig. \ref{fig:drl_horizon} illustrates the relationship between the optimization event horizon and the performance of the DRL model. The results indicate that shorter event optimization horizons, aligned with the average recovery rate, yield better outcomes. This suggests that the predictability of war dynamics, as described by the Lanchester model, allows the DRL model to focus on mitigating the more uncertain and impactful pandemic spread. Importantly, even with longer event optimization horizons, the model maintains a degree of robustness, though the influence of pandemic dynamics becomes more pronounced.

This study is not without limitations. First, the proposed model and simulation considered relatively short periods of time and therefore do not take into account population growth, reinfection, and multipathogen pandemics \cite{limit_1_5,limit_1_4,limit_1_3,limit_1_2,limit_1_1}. However, since wars can be extended over many years \cite{sullivan2008price}, this assumption may not hold. Future work could relax these assumptions to improve the model's robustness. In addition, the model assumes simplified military contact patterns and neglects the complexity of modern warfare. Although the current approach assumes a historical perspective in which battles occur far from civilian zones \cite{thorne2007battle}, extending the model to include civilian casualties common in contemporary conflicts \cite{khorram2021estimating} would improve its applicability. Furthermore, the model assumes isolation, ignoring international interdependencies during wars and pandemics \cite{limit_3_1}. Lastly, the simulation only considers one side of the war, neglecting the asymmetric nature of conflicts and assuming a static opposing force. Future studies should incorporate dynamic interactions between opposing forces to better capture real-world complexities \cite{sabin2012simulating}.

Taken together, our model provides valuable insights into designing effective healthcare and pandemic policies during wartime. For government agencies, understanding the predicted impact of conflict on disease spread can help prioritize resource allocation and intervention strategies. For military organizations, the model's predictions can guide operational planning to minimize health impacts on both civilian populations in war zones and military personnel. Further investigation is required to refine and extend the proposed model for greater practical utility.

\section*{Declarations}
\subsection*{Funding}
This study received no funding. 

\subsection*{Conflicts of interest/Competing interests}
None.

\subsection*{Code and Data availability}
The code and data that have been used in this study are publicly available in this study's GitHub repository: \url{https://github.com/shuchaa/pandemic_during_war_abs}.

\subsection*{Acknowledgments}
The authors wish to thank Noga Givon-Lavi for the clinical-epidemiological consulting.  

\subsection*{Author Contribution}
Adi Shuchami: Conceptualization, Formal Analysis, Investigation, Software, Visualization, Writing – original draft.\\
Teddy Lazebnik: Conceptualization, Data curation, Formal Analysis, Investigation, Methodology, Supervision, Visualization, Writing – original draft, Writing – review \& editing. \\  

\bibliography{biblio}
\bibliographystyle{unsrt}

\end{document}